\documentstyle[aps,prb,eqsecnum,epsf]{revtex}
\newcommand{\be}{\begin{equation}}
\newcommand{\ee}{\end{equation}}
\newcommand{\eea}{\end{eqnarray}}
\newcommand{\bea}{\begin{eqnarray}}
\newcommand{\nn}{\nonumber}

\newcommand{\lla}{\left\langle}     \newcommand{\rra}{\right\rangle}
\newcommand{\lbr}{\left(}           \newcommand{\rbr}{\right)}
\newcommand{\lsq}{\left[}           \newcommand{\rsq}{\right]}
\newcommand{\lcu}{\left\{}          \newcommand{\rcu}{\right\}}

\author{Kristian K.  M\"uller-Nedebock, Thomas A. Vilgis}
\address{Max-Planck-Institut f\"ur  Polymerforschung,
Postfach 3148, 55021 Mainz, Germany}
\title{Collective Dynamics of Random Polyampholytes}

\begin{document}

\maketitle


\begin{abstract}
We consider the Langevin dynamics of a semi-dilute system of chains which are
random polyampholytes of average monomer charge $q$ and with a
fluctuations in this charge of the size $Q^{-1}$ and with freely floating
counter-ions in the surrounding.  We cast the dynamics into
the functional integral formalism and average over the quenched charge
distribution in order to compute the dynamic structure factor and
the effective collective potential matrix.
The results are given for small charge fluctuations.  In the limit of finite
$q$ we then find that the scattering approaches the limit of
polyelectrolyte solutions.  
\end{abstract}
\pacs{PACS}


\section{Introduction}
\label{Introduction-Section}

The properties of polyampholytes are exceedingly complex and have 
recently been
the subject of various theoretical and simulational
investigations~\cite{JH,R,B1,B4,KK,stiff,Everaers,Efield}
ranging from  single-chain properties, the properties of solutions, 
to the absorption on walls.
The phase
behavior of these systems is rather rich and must be described by 
numerous parameters~\cite{Everaers}. When considering random,
quenched polyampholytes in statistical mechanics it is well known 
that the disorder requires the
implementation of of methods such as the replica trick~\cite{Mezard}.

In this paper we intend to draw upon previous work~\cite{Preprint} 
for polyelectrolytes to
describe these systems where there is random and quenched charge.  It is
notoriously difficult to deal with such systems in statistical
mechanics~\cite{Mezard}, by the necessity to introduce replicas.  However, as
has
been done for spin glasses~\cite{SompZipp} it is very convenient to formulate
the dynamics in the formalism of Martin, Siggia, and Rose and of Dominicis
and Peliti~\cite{MSR,DomPel}, to gain results.   Here we only 
restrict ourselves to
the case where the dynamics of the random polyampholyte chains are Rouse-like.
However, the system here differs from those of spin
glasses~\cite{Mezard,SompZipp} in that not the interaction but the charges
determining its sign and magnitude are the sources of randomness.  An average
over this quenched disorder as a rule then produces a more complex average for
the disorder than in the cases mentioned above.  Here we introduce such
approximations, and finally also compare the results to those for 
the dense polyelectrolyte solutions which has
been addressed in a similar formalism previously~\cite{Preprint}.

We introduce the model and its approximations in the next section using the
starting point of chains with charges uncorrelated along the backbones or
between chains.  We also retain the dynamics of pointlike ions in the fluid
in which the solution of polyampholyte chains is to be found.  
The dynamics are then transformed into a form with
collective variables after averaging
over the disorder (quenched, random charge distribution), with the use of the
random phase approximation (RPA), and by considering a specific limits of the 
distribution of the random charges.
The
structure factor and effective interactions are computed in the subsequent
section and the results compared with those of the polyelectrolytes.  The
contribution of the counter-ions to the effective potential is the familiar
Debye-H\"uckel form, originating in usage of a quadratic collective
approximation.  A further contribution comes from the randomness.  Since the
RPA is being implemented (Rouse modes are being used) we are not considering
here any collapsed or ``microgel-like'' phases, but work in the limit 
that the system is sufficiently dense to warrant the RPA.


\section{The Model}
\label{Model-Section}

The model consists of a number, $N_{p}$, of chains of equal length, $L$, which
behave according to the 
high friction regime of the Brownian dynamics.  The chains are modeled by the
Edwards Hamiltonian with Kuhn length $b$.
Associated with each segment (each arc position $s \in [0,L]$) 
of the chains is a quenched charge of magnitude $q_{i}(s) ds$ for the $i$th
random polyion chain, which is
assumed to have a Gaussian distribution, about an average charge density $q$.
Within this system there is
also a set of $N_{c}$ counter-ions, such that {\em on average} 
the random system is
electrically neutral.  Assuming that the counter-ions have a valence of one
this is expressed as,
\be
0= qN_{p}L + N_{c}.
\ee

Describing the random walks by variables
${\bf
  r}_{i}(s,t)$ and the counter-ions, which are assumed to be structureless, by
${\bf x}_{i}(t)$, the total energy of the system is given by the Hamiltonian,
\bea
H(t) & = & \frac{3}{2b^{2}}\sum_{i=1}^{N_{p}} \int_{0}^{L}ds\, \lbr
\frac{\partial {\bf r}_{i}(s,t)}{\partial s}\rbr^{2} + \sum_{i=1}^{N_{p}}\sum_{j=1}^{N_{p}}
\int_{0}^{L}\int_{0}^{L}ds\,ds'\,
\frac{2\pi \lambda_{B}q_{i}(s)q_{j}(s')}{|{\bf r}_{i}(s,t)- {\bf r}_{j}(s',t)|} \nn
\\
& & + \sum_{i=1}^{N_{p}}\sum_{j=1}^{N_{p}} \int_{0}^{L}\int_{0}^{L}ds\,ds'\,
v \delta\lbr {\bf r}_{i}(s,t)- {\bf r}_{j}(s',t)
\rbr + \sum_{i=1}^{N_{p}}\sum_{j=1}^{N_{c}}
\int_{0}^{L}ds\,
\frac{4\pi \lambda_{B}q_{i}(s)}{|{\bf r}_{i}(s,t)- {\bf x}_{j}(t)|}
\nonumber \\
& &
+ \sum_{i=1}^{N_{c}}\sum_{j=1,j\neq i}^{N_{c}}
\frac{2\pi \lambda_{B}}{|{\bf x}_{i}(t)- {\bf x}_{j}(t)|}.
\eea
The strength of the electrostatic interaction is given in terms of the Bjerrum
length,
\be
\lambda_{B} = \frac{e^{2}}{4\pi \epsilon k_{B}T},
\ee
where $e$ represents the electrical charge, $\epsilon$ the dielectric
constant, and  $k_{B}T=1/\beta$ the temperature multiplied by the Boltzmann
constant.  The polymer chains above can also subjected to an excluded volume
interaction of a strength given by $v$.  
The distribution of the charges $q_{i}(s)$ is assumed to be Gaussian for every
monomer on the chain independently:
\be
\label{charge}
P\lbr \lcu q_{i}(s) \rcu \rbr = \prod_{i=1}^{N_{p}}\prod_{s=0}^{L}\sqrt{ \frac{Q}{2\pi}} 
e^{-\frac{1}{2} Q (q_{i}(s) -q)^{2}}.
\ee
Here $Q$ indicates the magnitude of the fluctuations and $Lq$ represents an
average excess charge per macro-ion.  The product over arc-length of the
chains above is to be understood in a discretized sense.
Note that this means that even if the
parameter $q$ were considered to be zero, the chains would only be {\em
  statistically neutral}.

\subsection{Langevin Equations}
\label{Langevin-Section}

To start with, we follow the ideas of Fredrickson and 
Helfand in investigating the dynamics
of neutral polymers~\cite{FredricksonHelfand}.
The dynamics of the monomer positions of the two species of the 
system are now given by the appropriate
Langevin equations; for the random polyions these are:
\be
\label{Langevin1}
0 = {\cal L}_{ir}(s,t) = - \frac{\partial}{\partial t}{\bf r}_{i}(s,t)   
- \frac{1}{\zeta_{p}} \frac{\delta
  H(t)}{\delta {\bf r}_{i}(s,t)} + {\bf f}_{ir}(s,t).
\ee
The Langevin equations for the counter-ions are given by,
\be
\label{Langevin2}
0 = {\cal L}_{ix}(t) = - 
\frac{\partial}{\partial t}{\bf x}_{i}(t)   - \frac{1}{\zeta_{c}} \frac{\delta
  H(t)}{\delta {\bf x}_{i}(t)} + {\bf f}_{ix}(t).
\ee
The constants $\zeta_{p}$ and $\zeta_{c}$ are the friction coefficients of the
polyion segments and the counter-ions, respectively.  The expressions 
${\bf f}_{i,r}(s,t)$ and ${\bf f}_{i,x}(t)$ represent the random, Gaussian
noise exerted upon the individual particles.  
The averages of these forces are zero and the correlations are given by
the familiar expressions
$\lla {\bf f}_{i,x}(t) {\bf f}_{j,x}(t') \rra  =  2 k_{B}T\zeta^{-1}_{c}
\delta_{ij}\delta(t-t'){\bf 1}$  and $ \lla {\bf f}_{i,r}(s,t) {\bf f}_{j,r}(s',t')
\rra 
 =  2 k_{B}T\zeta^{-1}_{p}
\delta_{ij}\delta(t-t')\delta(s-s'){\bf 1}$.
There are $N_{p}$ and $N_{c}$
such equations, respectively.
For the sake of simplicity it is helpful to rescale the lengths above using 
a scale of temperature $k_{B}T=1$ and units of
length $b=\sqrt{3}$.

Now it is straightforward to implement the functional 
formalism of Martin, Siggia, and
Rose~\cite{MSR} (MSR), and of Dominicis and Peliti~\cite{DomPel},  in which the definition of the
functional delta function is used to convert the differential equations into a
functional form:
\bea
1 & = & \int \lcu\prod_{j=1}^{N_{p}} {\cal D}{\bf r}_{j}(s,t) 
{\cal D}{\bf \hat{r}}_{j}(s,t) \rcu \lcu \prod_{j=1}^{{N_{c}}} {\cal D}{\bf x}_{j}(t) 
{\cal D}{\bf \hat{x}}_{j}(t) \rcu J \times \nn \\
& & \times\exp \lsq i \sum_{j=1}^{N_{p}}
\int_{0}^{L} ds\int_{-\infty}^{+\infty} dt\, {\bf \hat{r}}_{j}(s,t) \cdot {\cal L}_{jr}(s,t)
+ i \sum_{j=1}^{N_{c}}\int_{0}^{L} ds\int_{-\infty}^{+\infty}dt\, 
{\bf \hat{x}}_{j}(t) \cdot {\cal L}_{jx}(t) \rsq, 
\label{MSR-f}
\eea
where $J$ is the Jacobian of the transformation.
Similarly, a generating functional can be defined schematically by
\bea
Z  \lsq\lcu{\bf h}_{j}(s,t)\rcu\rsq & = 
& \int \lcu\prod_{j=1}^{N_{p}} {\cal D}{\bf r}_{j}(s,t) 
{\cal D}{\bf \hat{r}}_{j}(s,t) \rcu \lcu \prod_{j=1}^{{N_{c}}} {\cal D}{\bf x}_{j}(t) 
{\cal D}{\bf \hat{x}}_{j}(t) \rcu J 
\nn \\ & & \times \exp \lsq i\sum_{j=1}^{N_{p}}\int_{0}^{L}ds\int_{-\infty}^{+\infty}dt\, 
{\bf h}_{j}(s,t)\cdot {\bf r}_{j}(s,t)\rsq\times \nn \\
& & \times\exp \lsq i \sum_{j=1}^{N_{p}}
\int_{0}^{L} ds\int_{-\infty}^{+\infty} dt\, {\bf \hat{r}}_{j}(s,t) \cdot {\cal L}_{jr}(s,t)
+ i \sum_{j=1}^{N_{c}}\int_{0}^{L} ds\int_{-\infty}^{+\infty}dt\, 
{\bf \hat{x}}_{j}(t) \cdot {\cal L}_{jx}(t)\rsq,
\eea
such that averages can be computed in terms of the functional derivatives with
respect to ${\bf h}$ of $Z$.  However, upon making use of causality
in taking the averages \cite{Jensen} the Jacobian can
be taken as being equal to 1.  This property of the MSR-formalism 
is especially useful when quenched disorder is present in the problem and it
has been applied previously to numerous problems, including the study of spin
glasses \cite{SompZipp} and of manifolds in a random potential
\cite{Vilgis,KinzelbachHorner}.   The rules based on causality entail that
any averages of products containing hatted fields ({\em i.e.} the field conjugate
to the Langevin expression) vanish if the latest time is found in the argument
of the hatted field.  If two times are equal, that of the hatted field is made
infinitesimally later to produce another vanishing average.

It is convenient to to introduce the following collective variables:
\bea
\rho_{jp}^{(1)}({\bf k}, s, t) & = & \exp \lsq -i{\bf k}\cdot{\bf r}_{j}(s,t)
\rsq
\nonumber\\
\rho_{jp}^{(2)}({\bf k}, s, t) & = & \frac{i{\bf k}\cdot{\bf
    \hat{r}}_{j}(s,t)}
{\zeta_{p}}\exp \lsq -i{\bf k}\cdot{\bf
  r}_{j}(s,t)\rsq \nonumber \\
\rho_{p}^{(1)}({\bf k}, s, t) & = & \sum_{j=1}^{N_{P}}\exp \lsq-i{\bf k}
\cdot{\bf r}_{j}(s,t) \rsq \nonumber \\
\rho_{p}^{(2)}({\bf k}, s, t) & = & \sum_{j=1}^{N_{P}} 
                \frac{i{\bf k} \cdot {\bf \hat{r}}_{j}(s,t)
}{\zeta_{p}}\exp \lsq -i{\bf k}\cdot{\bf r}_{j}(s,t) \rsq \nonumber \\
\rho_{c}^{(1)}({\bf k}, t) & = & \sum_{j=1}^{N_{c}}\exp \lsq -i{\bf k}
\cdot{\bf x}_{j}(t) \rsq \nonumber \\
\rho_{c}^{(2)}({\bf k}, t) & = & \sum_{j=1}^{N_{c}} 
                \frac{i{\bf k}\cdot{\bf \hat{x}}_{j}}{\zeta_{c}}\exp \lsq 
-i{\bf k}\cdot{\bf x}_{j}(t) \rsq
\eea
The collective variables with the superscript $(1)$ are the usual collective
densities of the polymer and the counterions in Fourier space.  The collective
variables with the superscripts $(2)$ refer to conjugate dynamical variables
of the density.
In the first two instances above these are simply density expressions for the
monomers.  The remaining definitions are those of collective variables for the
components of the system.
The MSR functional (\ref{MSR-f}) can now be rewritten in the form,
\be
1 = \int \lcu \prod_{j=1}^{N_{p}} {\cal D}{\bf r}_{j}(s,t) 
{\cal D}{\bf \hat{r}}_{j}(s,t) \rcu \lcu \prod_{j=1}^{{N_{c}}} {\cal D}{\bf x}_{j}(t) 
{\cal D}{\bf \hat{x}}_{j}(t) \rcu
\exp \lbr -i{\cal L}_{0}  -i{\cal L}_{1}\rbr,
\ee
where
\bea
-i{\cal L}_{0} & = &  i\sum_{j=1}^{N_{p}}\int_{0}^{L} ds
\int_{-\infty}^{+\infty}dt\, 
{\bf \hat{r}}_{j}(s,t) 
\cdot \lbr \frac{\partial {\bf r}_{j}(s,t) }{\partial t} + \frac{1}{\zeta_{p}}
\frac{\partial^{2} {\bf r}_{j}(s,t)}
{\partial s^{2}} + \frac{i}{\zeta_{p}}{\bf \hat{r}}_{j}(s,t)\rbr \nn \\
& & + i\sum_{j=1}^{N_{c}}\int_{-\infty}^{+\infty} dt\, {\bf \hat{x}}_{i}(t) 
\cdot \lbr \frac{\partial {\bf x}_{i}(t) }{\partial t} 
+ \frac{i}{\zeta_{c}}{\bf \hat{x}}_{j}(t)\rbr \\
-i{\cal L}_{1} & = & - \int_{-\infty}^{+\infty} dt\, \sum_{{\bf k}} \lbr {\underline
  \rho}_{c}^{T}({\bf k}, t) 
+ \sum_{j=1}^{N_{p}}\int_{0}^{L} ds \,
{\underline \rho}_{jp}^{T}({\bf k},s,t)q_{j}(s) \rbr \cdot {\bf V_{k}} \cdot
\nn \\
& & \cdot \lbr {\underline
  \rho}_{c}({\bf k}, t) 
+ \sum_{j'=1}^{N_{p}}\int_{0}^{L} ds' \,
{\underline \rho}_{j'p}({\bf k},s',t) q_{j'}(s') \rbr \nn \\ &&
- \int_{-\infty}^{\infty} dt\, \sum_{{\bf k}} \lbr {\underline
  \rho}_{p}^{T}({\bf k}, t) \rbr \cdot {\bf v} \cdot \lbr {\underline
  \rho}_{p}({\bf k}, t) \rbr
\eea
The sum over ${\bf k}$ is taken over all of Fourier space.
Here we have introduced the matrices with the two density variables,
\be
{\underline \rho} = \lbr \begin{array}{c} \rho^{{(1)}} \\ \rho^{(2)}
\end{array} \rbr
\ee
and the respective interaction matrices,
\be
{\bf V_{k}} = \lbr \begin{array}{cc} 0 & \frac{i\lambda_{B}}{k^{2}} \\ \frac{i\lambda_{B}}{k^{2}}
& 0 \end{array} \rbr
\ee
for the electrostatic potential, and for the excluded volume interaction:
\be
{\bf v} = \lbr \begin{array}{cc} 0 & iv \\ iv
& 0 \end{array} \rbr.
\ee
The appearance of the factor $i$ in the expressions concerning the potentials
above is purely a formality and comes from the fact that the dynamical formalism
is being used here.

It is now straightforward to implement the quenched charge average of
equation~(\ref{charge}).  In order to facilitate this we introduce a conjugate
vector field ${\underline \phi}$ \cite{NOTE} 
such that we can write the electrostatic part
of the equation as
\bea 
C\lbr \lcu q_{i}(s) \rcu \rbr & = & \exp \lsq
- \int_{-\infty}^{+\infty} dt\, \sum_{{\bf k}} \lbr {\underline
  \rho}_{c}^{T}({\bf k}, t) 
+ \sum_{j=1}^{N_{p}}\int_{0}^{L} ds \,
{\underline \rho}_{jp}^{T}({\bf k},s,t)q_{j}(s) \rbr \cdot {\bf V_{k}} \cdot\right.
\nn \\ & & 
 \left. \cdot \lbr {\underline
  \rho}_{c}({\bf k}, t) 
+ \sum_{j'=1}^{N_{p}}\int_{0}^{L} ds \,
{\underline \rho}_{j'p}({\bf k},s,t) q_{j'}(s) \rbr \rsq\nn \\ & = & 
{\cal N}_{\phi} \int {\cal D}{\underline \phi}({\bf k},t)
\exp \lsq -i\sum_{{\bf k}}\int_{-\infty}^{+\infty} dt\, {\underline \phi}^{T}({\bf k},t)
\cdot \lbr {\underline
  \rho}_{c}^{T}({\bf k}, t) 
+ \sum_{i}^{N_{p}}\int ds 
{\underline \rho}_{ip}^{T}({\bf k},s,t)q_{i}(s) \rbr\rsq \nn \\ & & \times
\exp \lsq -\sum_{k}\int_{-\infty}^{+\infty} dt\, {\underline \phi}^{T}({\bf k},t)
\cdot {\bf V_{k}}^{{-1}} \cdot {\underline \phi}({-{\bf k},t}) \rsq
\label{e228}.
\eea
The normalization of the Gaussian integration above is given by the factor
${\cal N}_{\phi}$.
Consequently, we are now in a position to take an average with respect to the
quenched disorder of the expression above using eq.~(\ref{charge}).  The fact
that the MSR-formalism has a normalization equal to one permits the evaluation
of this quenched average without the need to resort to the replica trick.
Denoting the $q$-averaged quantities [having used the 
distribution (\ref{charge})]
by an overbar the following result is imminent:
\bea
{\overline C} & = & -\sum_{{\bf k}}\int_{-\infty}^{+\infty}dt\, {\underline
  \phi}^{T}({\bf k},t)\cdot{\bf
  V_{k}}^{-1}\cdot {\underline \phi}(-{\bf k},t) + i \sum_{{\bf k}} 
\int_{-\infty}^{+\infty}dt\,
{\underline
  \phi}^{T}({\bf k},t) \cdot \lsq q {\underline \rho}_{p}(-{\bf k},t) 
+ {\underline
  \rho}_{c}(-{\bf k},t) \rsq \nn \\
& & - \sum_{{\bf k}}\sum_{{\bf k}'} \int_{-\infty}^{+\infty}dt
\int_{-\infty}^{+\infty}dt'\,
{\underline \phi}^{T}({\bf k},t)\cdot \sum_{j=1}^{N_{p}}\int_{{0}}^{L}ds\,\lbr
{\underline \rho}_{jp}(-{\bf k},s,t) Q^{-1} {\underline \rho}_{jp}({\bf
  k'},s,t') \rbr \cdot {\underline \phi}(-{\bf k},t).
\eea
We see that the effect of disorder is to alter the interaction between the
components of the system, by a term dependent on the configuration of the
system itself.  This is familiar from averages over Gaussian distributions
of disorder.
Integrating over the fields ${\underline \phi}$ leads to the disorder-averaged
expression of the MSR generating functional such that the electrostatic part of
${\cal L}_{1}$ becomes replaced by,
\bea
\label{average}
\ln {\overline{\exp-i{ {\cal L}_{1}}}} & = &  -\frac{1}{2} {\rm Tr} \ln \lsq 
{\bf V}^{-1}({\bf k})\delta({\bf k}-{\bf k}') +
{\bf W}({\bf k},{\bf k}')    Q^{-1} \rsq \nn \\
& &  - \sum_{{\bf k}}\sum_{{\bf k}'} 
\int_{-\infty}^{+\infty}dt\int_{-\infty}^{+\infty}dt'\,
\lsq q {\underline \rho}_{p}^{T}(-{\bf k},t) + {\underline
  \rho}_{c}^{T}(-{\bf k},t) \rsq\cdot \nonumber \\
& & \cdot \lbr {\bf V}^{-1}({\bf k})\delta({\bf k}-{\bf k}') +
{\bf W}({\bf k},{\bf k}')Q^{-1}   \rbr^{-1} 
\cdot \lsq q {\underline \rho}_{p}(-{\bf k}',t') 
+ {\underline
  \rho}_{c}(-{\bf k}',t') \rsq.
\eea
Here the trace (Tr) is taken over the $({\bf k},{\bf k}')$--space, 
over the times and the density
matrix indices.  The definition of the matrix ${\bf W}$ is,
\be
{\bf W}({\bf k},{\bf k}') = \sum_{j=1}^{N_{p}}\int_{0}^{L}ds\,\lbr
{\underline \rho}_{jp}(-{\bf k},s,t)  {\underline \rho}_{jp}^{T}({\bf
  k'},s,t') \rbr .
\ee
It is always associated with a factor of $1/Q$ and it can be seen that this
matrix is proportional to the mass density of polyampholyte chains in the system.
This expression (\ref{average}) together with the remaining terms
of the system given earlier, and which are still contained in ${\cal L}_{0}$
is 
exact, but must be approximated in order to make further progress.

\subsection{Approximation for small $Q^{-1}$ and finite $q$}
\label{Approximation-Section}

In order to be able to approximate further the expression above and to compute
the dynamical structure factor of the present system of polyampholytes we
shall assume the fluctuations of the quenched charges to be small, such that
we can expand the determinant and invert the 
matrix of eq.~(\ref{average}) to first
order in $Q^{-1}$.

Dealing first with the determinant we see that
\bea
\exp \lbr{ -\frac{1}{2} {\rm Tr} \ln \lbr {\bf V}^{-1} + Q^{-1}{\bf W} \rbr}\rbr 
& = &
\exp\lbr{ -\frac{1}{2} {\rm Tr} \ln  {\bf V}^{-1}}\rbr
\exp\lbr{ -\frac{1}{2} {\rm Tr} \ln \lbr {\bf 1} + Q^{-1}{\bf V}{\bf W} \rbr}\rbr \nn \\
& \simeq & \exp\lbr{ -\frac{1}{2} {\rm Tr} \ln  {\bf V}^{-1}} \rbr
\exp\lbr{-\frac{1}{2} {\rm Tr} Q^{-1} \lbr i\frac{\lambda_{B}}{k^{2}}\rbr {\bf M}
 + {\cal O}(Q^{-2})}\rbr \\
& = & \exp\lbr{ -\frac{1}{2} {\rm Tr} \ln  {\bf V}^{-1} +  {\cal O}(Q^{-2})     }\rbr.
\eea
The above matrix, ${\bf M}$, is defined by:
\be
{\bf M} = \lbr
\begin{array}{cc}   
{ \rho}_{jp}^{(2)}(-{\bf k},s,t)  { \rho}_{jp}^{(1)}({\bf k'},s,t') &
{ \rho}_{jp}^{(2)}(-{\bf k},s,t)  { \rho}_{jp}^{(2)}({\bf k'},s,t') \\
{ \rho}_{jp}^{(1)}(-{\bf k},s,t)  { \rho}_{jp}^{(1)}({\bf k'},s,t') &
{ \rho}_{jp}^{(1)}(-{\bf k},s,t)  { \rho}_{jp}^{(2)}({\bf k'},s,t')
\end{array} \rbr.
\ee
The result follows from the isotropic ${\bf k}$--integration (note 
the definitions of the collective variables) and the
off-diagonal property of the interaction matrix.  We have used the fact that
the trace also entails taking the summations with $\delta({\bf k}- {\bf k'})$
and $\delta ( t -t')$.

The matrix $({\bf V}^{-1} + {\bf W}Q^{-1})^{-1}$ sandwiched between the two
collective densities in equation~(\ref{average}) is treated also by an expansion:
\bea
\lbr {\bf V}^{-1} + {\bf W}Q^{-1}\rbr^{-1} & = & {\bf V} \lbr {\bf 1} +
Q^{-1}{\bf W\cdot V} \rbr^{-1} \nn \\
& = & {\bf V} \lbr {\bf 1} -
Q^{-1}{\bf W\cdot V} +\ldots \rbr.
\eea
The second term above is analogous to the 
term sometimes denoted by ``polyampholyte
attraction''~\cite{JH}. 
By using the definitions of the matrices involved the second term of the 
expansion above looks
as follows:
\bea
-Q^{-1} {\bf V} \cdot {\bf W} \cdot {\bf V} &  = &
\frac{Q^{-1}\lambda^{2}_{B}}{k^{2}(k')
^{2}} \sum_{j=1}^{{N_{p}}}\int_{0}^{L}ds\,
\lbr \begin{array}{cc} 0 & 1\\ 1& 0 \end{array}\rbr \times \nn \\
& & \times \lbr
\begin{array}{cc} \rho_{jp}^{(1)}(-{\bf k},t,s)\rho_{jp}^{(1)}({\bf k}',t',s) 
& \rho_{jp}^{(1)}(-{\bf k},t,s)\rho_{jp}^{(2)}({\bf k}',t',s)
\\ \rho_{jp}^{(2)}(-{\bf k},t,s)\rho_{jp}^{(1)}({\bf k}',t',s) &
\rho_{jp}^{(2)}(-{\bf k},t,s)\rho_{jp}^{(2)}({\bf k}',t',s) 
\end{array}\rbr 
\lbr\begin{array}{cc} 0 & 1\\ 1& 0 \end{array}\rbr \nn \\
& = &  
\frac{Q^{-1}\lambda^{2}_{B}}{k^{2}(k')
^{2}} \sum_{j=1}^{{N_{p}}}\int_{0}^{L}ds\,\lbr
\begin{array}{cc} \rho_{jp}^{(2)}(-{\bf k},t,s)\rho_{jp}^{(2)}({\bf k}',t',s) 
& \rho_{jp}^{(2)}(-{\bf k},t,s)\rho_{jp}^{(1)}({\bf k}',t',s)
\\ \rho_{jp}^{(1)}(-{\bf k},t,s)\rho_{jp}^{(2)}({\bf k}',t',s) &
\rho_{jp}^{(1)}(-{\bf k},t,s)\rho_{jp}^{(1)}({\bf k}',t',s) 
\end{array}\rbr
.
\label{second-term}
\eea
This matrix is non-diagonal in the time and Fourier component variables.
However, it is totally diagonal in the monomer labels, which follows from the
uncorrelated nature of the distribution of charges.  Introducing a correlation
would be a straightforward generalization of our methods.

We wish to implement the random phase approximation in the 
collective variables and deal with
contributions in the exponent up to quadratic order in the MSR functional.
Consequently, one of the simplest approximations entails the substitution of the
average of the expression (\ref{second-term})
above into the equation (\ref{average}) to create an effective
potential between the collective variables.  This approximation is best when
the polyampholyte chains have small fluctuations and retain a more-or-less
Gaussian distribution in the dense, yet unentangled, system considered
here.  For these
purposes we consider polyions of which the excess charges are not too small to
describe the collapsed regime.  The collapsed regime must be descibed in a
completely different formalism (see, {\em e.g.} \cite{Barbosa1}).

Let the following matrix be defined for an arbitrary $j\in \lcu
1,\ldots,N_{p}\rcu$: 
\bea 
{\bf w} & = &
\lbr \begin{array}{cc} 0 & w_{12}({\bf k},\omega) \\ w_{21}({\bf k},\omega) &
  w_{22}({\bf k},\omega) \end{array}\rbr \nonumber \\ & = & 
\frac{1}{2\pi}\int_{-\infty}^{+\infty} \int_{-\infty}^{+\infty}
dt dt'\, e^{{i\omega (t-t')}} N_{p} \int_{0}^{L} ds\, \lla \lbr 
\begin{array}{cc} \rho_{jp}^{(2)}(-{\bf k},t,s)\rho_{jp}^{(2)}({\bf k}',t',s) 
& \rho_{jp}^{(2)}(-{\bf k},t,s)\rho_{jp}^{(1)}({\bf k}',t',s)
\\ \rho_{jp}^{(1)}(-{\bf k},t,s)\rho_{jp}^{(2)}({\bf k}',t',s) &
\rho_{jp}^{(1)}(-{\bf k},t,s)\rho_{jp}^{(1)}({\bf k}',t',s) 
\end{array}\rbr \rra_{0} .
\label{eqn-w}
\eea
We shall use this matrix, where averaging has taken place using the
distribution given with ${\cal L}_{0}$ in approximation [see equation 
(\ref{av-0})].

The computation is continued by the transformation to a random phase
approximation (RPA) for the collective variables introduced earlier.  
By denoting the appropriate
average of a functional ${\sf Q}$ by
\be
\lla {\sf Q} \rra_{0} = {\cal N} \int \lcu \prod_{j=1}^{N_{p}} {\cal D} {\bf r}_{j} 
{\cal D}{\bf \hat{r}}_{j} \rcu
\lcu \prod_{j=1}^{N_{c}} {\cal D} {\bf x}_{j} {\cal D}{\bf \hat{x}}_{j}\rcu {\sf Q} 
e^{-i{\cal L}_{0}},
\label{av-0}
\ee
such that the integration for the structure factor and response functions
takes place over the collective variables ${\underline \rho}$ with mean $\lla
{\underline \rho}\rra$ and distribution $\exp -\frac{1}{2}{\underline \rho}\cdot {\bf
  S}_{0}^{-1} \cdot 
{\underline \rho}$, where
\be
{\bf S}_{0} = \lla {\underline \rho} ({\bf k},\omega) {\underline \rho} 
({\bf k}',\omega')
\rra_{0}.
\ee
The derivations of these quantities are straightforward and follow exactly
along the lines of the paper of Fredrickson and 
Helfand~\cite{FredricksonHelfand}.  The results are summarised in the
Appendix. 
Furthermore, at very large length-scales it can be assumed that the elements
of the matrix ${\bf w}$ are proportional the related structure factor.


\section{Dynamic Structure Factor}
\label{Structure-Factor-Section}

From the work in the previous section we are now in a position to write down
the expression for dynamic collective averaging.  This is given by:
\be
\lla {\sf Q} \rra_{coll} = {\cal N} \int {\cal D}{\underline \rho}_{c}
{\cal D}{\underline \rho}_{p} {\sf Q}
\exp \lbr - {\underline \rho}_{c}( {\bf S}_{0,c}^{-1} + {\bf A}) {\underline
  \rho}_{c} -
{\underline \rho}_{p}( {\bf S}_{0,p}^{-1} +  {\bf E} + q^{2} {\bf A}) {\underline \rho}_{p}
- {\underline \rho}_{c} q {\bf A} {\underline \rho}_{p} \rbr.
\ee
Here ${\cal N}$ is the appropriate normalizing factor.
The matrix ${\bf A}$ has been defined through the approximations above as,
\be
{\bf A} = {\bf V} + \frac{Q^{-1}\lambda_{B}^{2}}{k^{4}} {\bf w}.
\ee
Integration over the counter-ion collective coordinates now gives the
result,
\bea
\lla {\sf Q} \rra_{coll} & = & {\cal N} \int
{\cal D}{\underline \rho}_{p} {\sf Q}
\exp \lbr -
{\underline \rho}_{p}( {\bf S}_{0,p}^{-1} +  {\bf v} + q^{2} {\bf A} 
- q^{2} {\bf A} ({\bf S}_{0c}^{-1}+ {\bf A})^{-1}{\bf A}  )  {\underline \rho}_{p}
 \rbr \nn \\
& = & {\cal N} \int
{\cal D}{\underline \rho}_{p} {\sf Q}
\exp \lbr -
{\underline \rho}_{p}( {\bf S}_{0,p}^{-1} + {\bf B} ) {\underline \rho}_{p}
\rbr,
\eea
with ${\bf B} = {\bf v} + q^{2} {\bf A} 
- q^{2} {\bf A} ({\bf S}_{0c}^{-1}+ {\bf A})^{-1}{\bf A}$.
The (1,1)-component of the inverse of the expression above is the dynamical
structure factor.  By denoting the interaction term above as the matrix ${\bf
  B}$
the structure factor $G$ has the form:
\be
\label{StructFact}
G = \frac{S_{0p,11} - B_{22}S_{0p,12}S_{0p,21}}{( 1 + B_{21}S_{0p,12})( 1 + B_{12}S_{0p,21})}.
\ee
We point out that it is less tedious to compute the structure factor by taking
the RPA average over the counter-ions before the integration over ${\underline \phi}$.

\subsection{Effective Interaction}

The above form can be made to look even far simpler by noticing that it is
convenient to perform the average over the counter-ions in eq. (\ref{e228})
already.  The potential expression ${\bf V_{k}}$ then becomes replaced by
${\bf V}^{\rm eff}_{k\omega}$ which contains the screening, and the matrix
${\bf B}$ above becomes simply,
\be
{\bf B}={\bf v} + q^{2}{\bf V}^{\rm eff} - q^{2}{Q^{-1}}{\bf V}^{{\rm eff}}\cdot {\bf
  W}\cdot {\bf V}^{\rm eff}.
\label{interacteq}
\ee
It represents the effective interactions between the poly-ionic collective
variables after averaging over the disorder.  The term ${\bf V}^{{\rm eff}}$
represents the effective interchain and intrachain interaction Coulomb
interaction before the averaging over the quenched charges on the
polyampholyte chains, after integration over the free counter-ion (collective)
degrees of freedom.
We note the typical signature of the random quenched polyampholyte 
interaction~\cite{JH}
in the appearance of a term $\propto k^{{-4}}$ and proportional to $Q^{-1}$.
  
For $R_{g}k \ll 1$ it is simple to
express the matrices [where ${\bf w}$ is related to ${\bf W}$ as in
eq.~(\ref{eqn-w})] 
\be
{\bf w} L \simeq \lbr \begin{array}{cc} 0 & \rho_{p} \frac{k^{2}/\zeta_{p}}{\omega+
    ik^{2}/L\zeta_{p}}\\
\rho_{p} \frac{k^{2}/\zeta_{p}}{-\omega+
    ik^{2}/L\zeta_{p}} & \rho_{p}\frac{2k^{2}/\zeta_{p}}{\omega^{2} + \lbr
    k^{2}/L\zeta_{p} \rbr^{2} }\end{array}\rbr
\ee
and
\be
{\bf V}^{{\rm eff}} = \lbr \begin{array}{cc} 0 &
  \frac{i\lambda_{B}q^{2}}{k^{2}}
\frac{ k^{2} -i\omega\zeta_{c} }{ k^{2} + \kappa^{2} -i\omega\zeta_{c}  } \\
\frac{i\lambda_{B}q^{2}}{k^{2}}
\frac{ k^{2} +i\omega\zeta_{c} }{ k^{2} + \kappa^{2} +i\omega\zeta_{c}  } & 
\frac{ 2 \lambda_{B}^{2}q^{2}\zeta_{c} \rho_{c} }{ (k^{2} + \kappa^{2})^{2}  +
  (\zeta_{c}\omega)^{2} } \end{array} \rbr.
\ee
The effective Coulomb interaction as above is the same is found in the case of
polyelectrolytes~\cite{Preprint}
with the charge density along the chain replaced by $q$.  It is easy to see
that the off-diagonal terms above in the limit $\omega \rightarrow 0$ return
the familiar Debye-H\"uckel screening potential.  The formalism of
coupled, dynamical equations leads to the 
additional diagonal component in the matrix above,
which would not have resulted had the counter-ion dynamics been neglected and
a
Debye-H\"uckel potential used from the start.
The definition of the Debye screening parameter is given by
\be
\kappa^{2} = \lambda_{B} \rho_{c}
\label{kappaeq}
\ee
as usual.

\subsection{Structure Factor}

By substitution of all the above into the expression for the structure factor,
eq.~(\ref{StructFact}), we are now in a position to 
discuss its properties.   
We emphasize again that in the approximations used, terms up to order $1/Q$
can only be considered, where the effect of the term $Q^{-1}q^{2}V_{\rm
  eff}WV_{\rm eff} \propto {\cal O} (q^{2}\lambda_{B}^{2}/(Q\kappa^{2}))$.
We plot the typical structure factor in Figure 1.
 A peak is noticed, as for polyelectrolytes and
in in the limit as
$Q\rightarrow \infty$ and the results for polyelectrolytes are 
and must be recovered.  For extremely small scattering vector ${\bf k}$ and
frequency $\omega$ a behavior of the structure factor proportional to
$k^{2}/\omega^{2}$ is expected as shown in \cite{FredricksonHelfand}.

The general properties of the peak depend on the parameters $q, \kappa,
\lambda_{B}, v$, the polyampholyte density.  Firstly, it is noticed that the
average charge fraction $q$ per monomer when increased causes a decrease
in the peak height.  This is plotted in Figure 2.  Figure 3 shows that an
increase in the screening parameter, or a corresponding increase of salt (see
eq. (\ref{kappaeq})) concentration of the system causes an increase of the
peak,
whereas a decrease of the maximal height is the result of turning on the
excluded volume potential for the chain (Figure 4).  These properties are in
general agreement with those already investigated for polyelectrolytes
\cite{Preprint,VilgisBorsali}.

Secondly, the role of the quenched fluctuations of polyampholyte charge, which
are contained within the parameter $Q$ auch that 
$(q^{2}\lambda_{B}^{2}/(Q\kappa^{2}))$ is small
need to be considered.  
By comparing with equation (\ref{interacteq}) it is clear that the
fluctuations-related term is proportional the Bjerrum length squared in
distinction to the other term of the interaction.
Consequently in Figure 5 we give a plot of three curves of the dynamical
structure fact at finite frequency $\omega$, for a value of $Q=10$.  It is
seen that an increase of the Bjerrum length leads to a decrease of the maximal
height.  In order to understand the magnitude of the role of the fluctuations
we now compare the polyampholyte (with $Q=10$) to a system which is almost
completely a polyelectrolyte ($Q=10^{6}$) in Figure 6.  A charge density of
$q=0.01$ has been assumed in these plots of the maximum at 
frequency $\omega=0.05$ in dependence of $\lambda_{\rm B}$, with $\kappa=1$.
The higher peak of the polyampholyte in distinction to the polyelectrolyte (of
the same backbone charge density) is clearly shown, although the changes in
the regime where our approximations are most valid are small.  

We note that if the Bjerrum length were to be increased further, the
criterion for counterion condensation 
$\lambda_{B}q \simeq 1$
would be achieved and our gaussian model
would lose its applicability.



\section{Discussion}
\label{Discussion-Section}

We have briefly shown how the formalism of Fredrickson and Helfand can be used
to describe the collective dynamics of randomly charged chains, which are
still assumed to be in the Gaussian regime.  The structure factor shows a
marked peak at small frequencies and low wavevector dependent on the charge
fluctuation strength $Q$ this peak can be understood to emerge from the mainly
polyelectrolyte behavior of our system.  The effects of the small
polyampholyte quenched charge fluctuations are in increasing the height of the
dynamical polyelectrolyte peak.  Here the dependence upon the Bjerrum length
are particularly evident.

Clearly, the approximations we have made to the very lowest orders are
amenable to improvement, and it is true that a self-consistent treatment of
the problem of polyampholytes will provide a better picture.  This is
especially so in respect to the Gaussian chain assumption which has gone into
the computation of our collective (non-interacting) distribution.  Since
polyampholyte conformations clearly occur in many non-gaussian (e.g. collapsed)
phases~\cite{JH,R,Barbosa1}, 
the altered single chain properties are the future topic
of our investigations.

\section*{Acknowledgement}

The funding of this work  by the Deutsche Forschungsgemeinschaft:
Schwerpunkt Polyelektrolyte is most gratefully acknowledged.

\appendix

\section*{RPA Results}

The RPA has been computed before by Fredrickson and
Helfand~\cite{FredricksonHelfand} and we give the 
results for the conditions
$kR_{g} \ll 1$:
\bea
\lla \rho_{p}^{(1)}(-{\bf k},-\omega) \rho_{p}^{(1)}({\bf k}, 
\omega) \rra_{0} & = 
& \frac{2k^{2}\zeta_{p}}{\omega^{2}\zeta_{p}^{2} + k^{4}L^{-2}} \\
\lla \rho_{c}^{(1)}(-{\bf k},-\omega) \rho_{c}^{(1)}({\bf k}, 
\omega) \rra_{0} & = 
& \frac{2k^{2}\zeta_{c}}{\omega^{2}\zeta_{c}^{2} + k^{4}} \\
\lla \rho_{p}^{(1)}(-{\bf k},-\omega) {\bf k}\cdot
\rho_{p}^{(2)}({\bf k}, \omega) \rra_{0} & = 
& \frac{2k^{2}}{\omega\zeta_{p} + ik^{2}L^{-1}} \\
\lla \rho_{c}^{(1)}(-{\bf k},-\omega) {\bf k}\cdot
\rho_{c}^{(2)}({\bf k}, \omega) \rra_{0} & = 
& \frac{2k^{2}}{\omega\zeta_{2} + ik^{2}}.
\eea
All remaining
pairwise correlations are zero.



\newpage

\begin{figure}
\epsfxsize 10cm
\epsffile{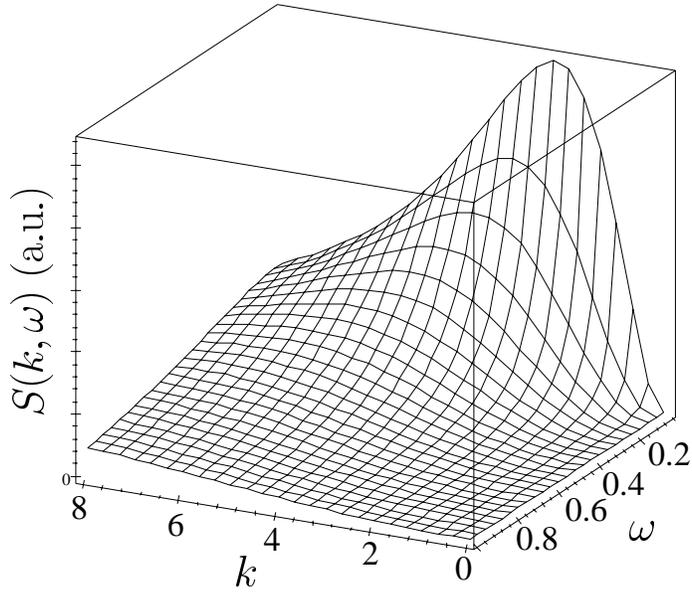}
\caption{Plot of the dynamical structure factor of the polyampholyte solution
  case.  Here $Q=5$, the friction coefficients equal 1, the Bjerrum length is
  set to 1 and the chain length 100, the average charge fraction 0.2,
  the Debye length equals 1, and there is no excluded volume potential.  The
  vertical units are arbitrary.}
\end{figure}
\newpage
\begin{figure}
\epsfxsize 10cm
\epsffile{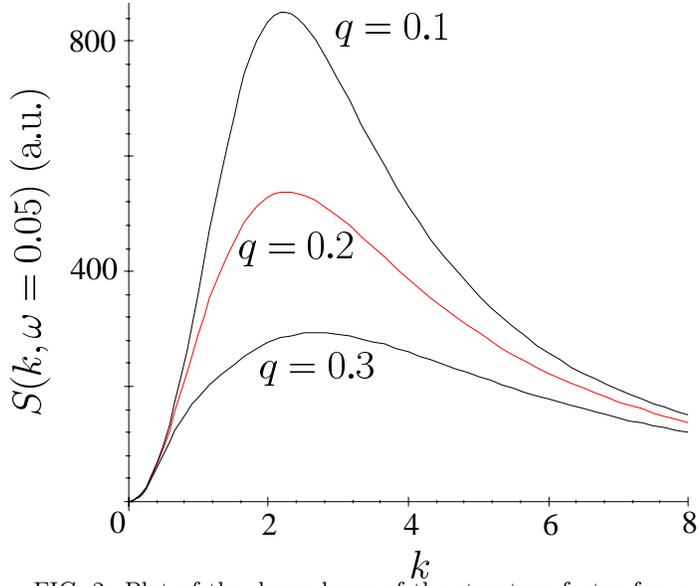}
\caption{Plot of the dependence of the structure factor for $\omega=0.05$,
$v=0$, $\lambda_{\rm B}=1$, $Q=5$, $ \kappa=1$, and
$ L=100$ for three different values of the
average monomer charge $q$.}
\end{figure}

\begin{figure}
\epsfxsize 10cm
\epsffile{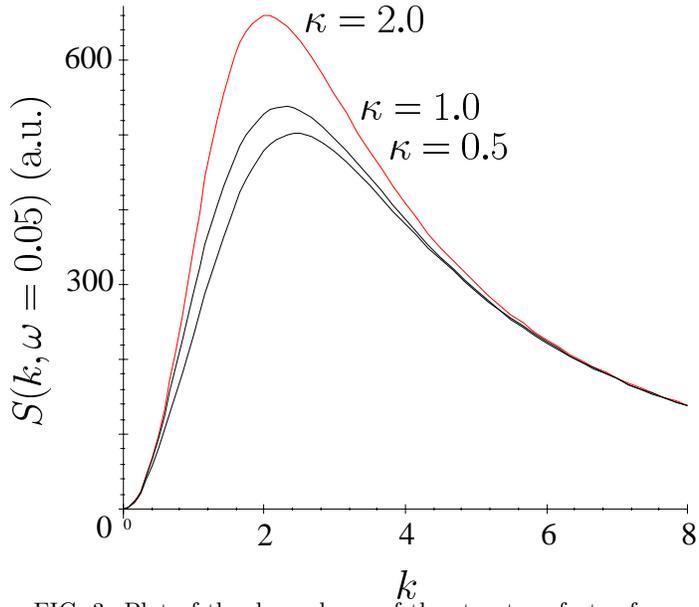}
\caption{Plot of the dependence of the structure factor for $\omega=0.05$,
$v=0$, $\lambda_{\rm B}=1$, $Q=5$, $ q=0.2$, and  $L=100$ for three 
different values of the
parameter $\kappa$.}
\end{figure}

\begin{figure}
\epsfxsize 10cm
\epsffile{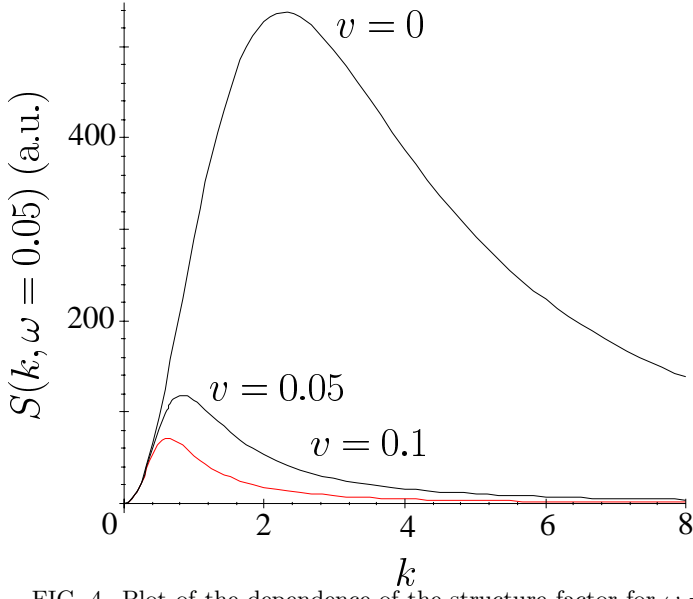}
\caption{Plot of the dependence of the structure factor for $\omega=0.05$,
$q=0.2$, $\lambda_{\rm B}=1$, $Q=5$, $\kappa=1$, and $L=100$ for three 
different values of the
excluded volume interaction strength.}
\end{figure}

\begin{figure}
\epsfxsize 10cm
\epsffile{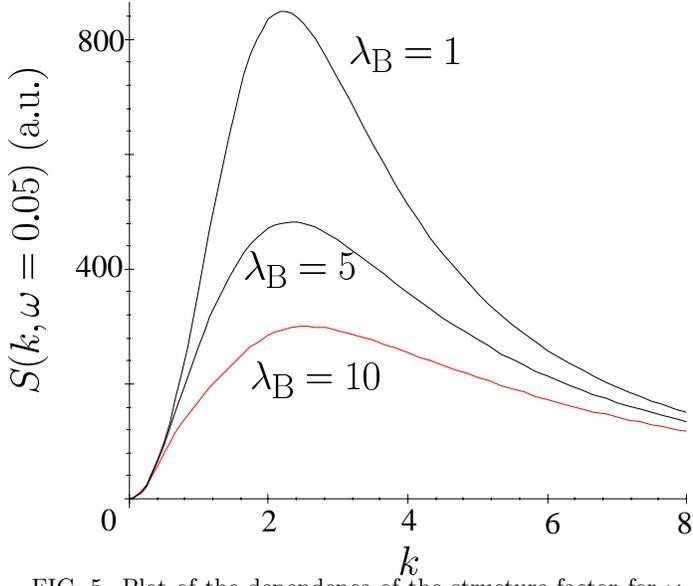}
\caption{Plot of the dependence of the structure factor for $\omega=0.05$,
$v=0$, $q=0.2$, $Q=10$, $\kappa=1$, and $L=100$ for three different values of the
Bjerrum length.}
\end{figure}

\begin{figure}
\epsfxsize 10cm
\epsffile{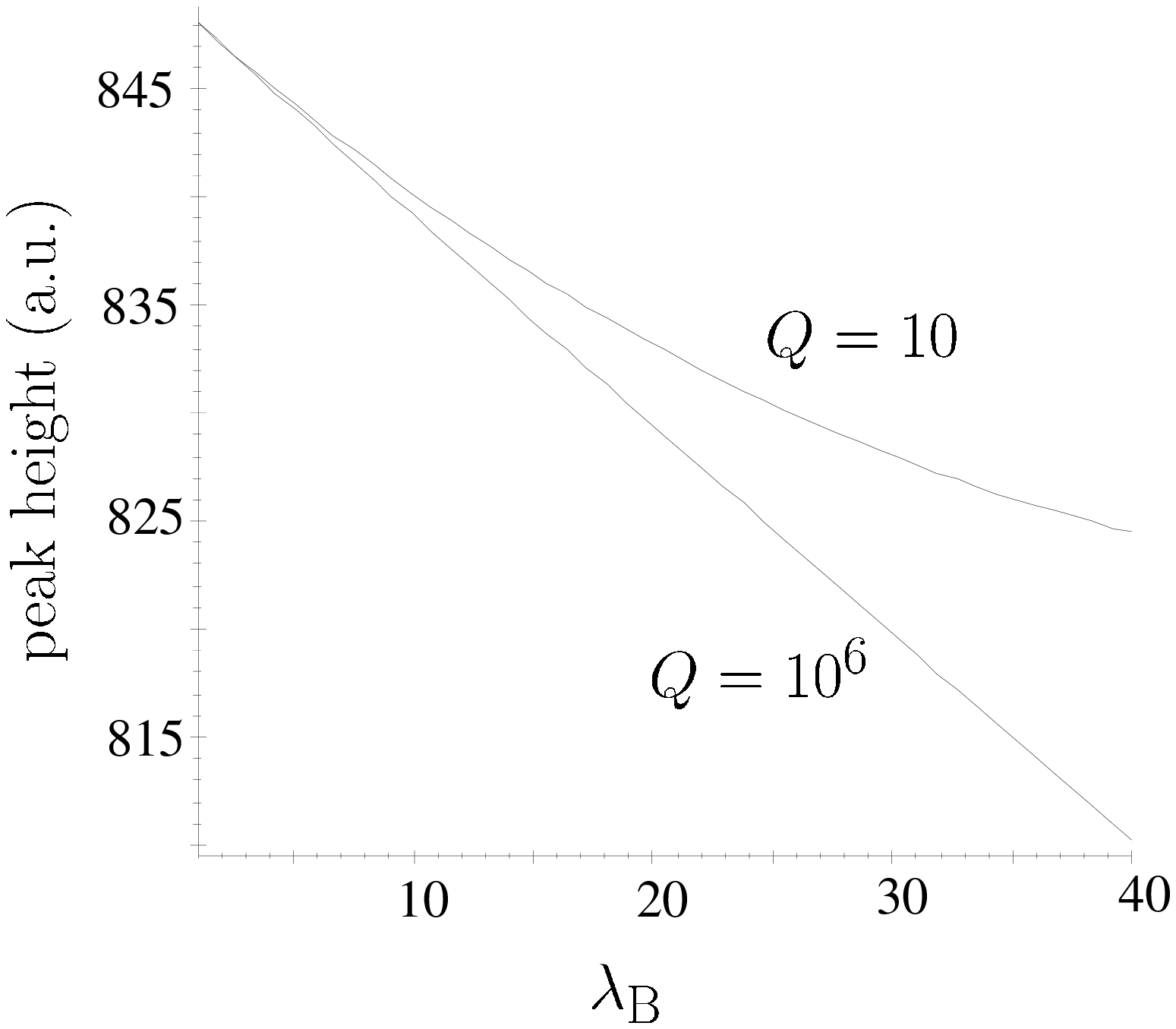}
\caption{Plot of the peak height at $\omega=0.05$,
$v=0$, $\lambda_{\rm B}=1$, $q=0.01$, $\kappa=1$, and $L=100$ 
for the cases $Q=10$ and $Q=10^{6}$.}
\end{figure}


\begin{thebibliography}{99}
\bibitem{NOTE} We remark that the alternative approach of introducing the
  polyampholyte charge density through a paramatrized delta function, produces
  the same results as in this formal approach.
\end{thebibliography}
\end{document}